\documentclass[aps,prb,twocolumn,superscriptaddress,showpacs,amsmath,amssymb]{revtex4}

\usepackage{amsfonts}
\usepackage{amssymb,amsmath}
\usepackage{graphicx}
\usepackage{dcolumn}
\usepackage{bm}

\begin{document}

\title{Low-temperature thermodynamics
       for a flat-band ferromagnet:
       Rigorous versus numerical results}

\author{Oleg Derzhko}
\affiliation{Institute for Condensed Matter Physics,
          National Academy of Sciences of Ukraine,
          L'viv-11, 79011, Ukraine}
\affiliation{Institut f\"ur Theoretische Physik,
          Universit\"at Magdeburg,
          P.O. Box 4120, 39016 Magdeburg, Germany}
\author{Andreas Honecker}
\affiliation{Institut f\"ur Theoretische Physik,
          Georg-August-Universit\"at G\"ottingen,
          37077 G\"ottingen, Germany}
\author{Johannes Richter}
\affiliation{Institut f\"ur Theoretische Physik,
          Universit\"at Magdeburg,
          P.O. Box 4120, 39016 Magdeburg, Germany}

\date{August 27, 2007}


\begin{abstract}
The repulsive Hubbard model on a sawtooth chain exhibits
a lowest single-electron band which is completely dispersionless (flat)
for a specific choice of the hopping parameters.
We construct exact many-electron ground states for electron fillings
up to 1/4.
We map the low-energy degrees of freedom of the electron model
to a model of classical hard dimers on a chain
and, as a result,
obtain the ground-state degeneracy as well as
closed-form expressions for the low-temperature
thermodynamic quantities around a particular value of the chemical potential.
We compare our analytical findings
with complementary numerical data.
Although we consider a specific model,
we believe that some of our results such as a low-temperature
peak in the specific heat are generic
for flat-band ferromagnets.
\end{abstract}

\maketitle

{\it Introduction and motivation.}
The Hubbard model is the simplest model for strongly interacting
electrons in a solid.\cite{thehubbard}
Nevertheless, the analysis of its properties is a difficult task.
Only relatively few rigorous results are known,
see e.g.\ Refs.~\onlinecite{thehubbard,lieb,tasaki_jpcm}.
A special focus of rigorous studies
is the search for ground state (GS) ferromagnetism (FM).
\cite{vollhardt,mielke,tasaki,tasaki_ptp}
In particular,
different (nonbipartite) lattices
supporting dispersionless (flat) single-electron bands
were studied in some detail.
The Hubbard model on such lattices may have
ferromagnetic GS's for
certain electron concentrations
(so-called flat-band FM).\cite{mielke,tasaki,tasaki_ptp}
An important aspect of these studies is the existence of eigenstates
where electrons are localized on finite areas of the lattice.
Recently, the concept of flat-band FM has been developed further,
see e.g.\ Refs.~\onlinecite{ichimura,tanaka,sekizawa,nishino,tanaka_tasaki},
and relations to materials with (almost) flat bands have been pointed out.
Another recent development
is the discovery of a class  of exact eigenstates
for antiferromagnetic spin models on certain frustrated lattices.
\cite{lm1}
These states,
called localized-magnon states,
are GS's of the $XXZ$ Heisenberg antiferromagnet (AFM) in strong
magnetic fields,
and lead to interesting low-temperature physics
near the saturation field
such as
a macroscopic magnetization jump,\cite{lm1}
a field-tuned lattice instability,\cite{lm5}
a finite residual entropy,\cite{lm6a,lm6,lm7}
an enhanced magnetocaloric effect,\cite{lm6a}
or a finite-temperature order-disorder phase transition
in
2D
Heisenberg spin systems.
\cite{lm7}
On the one-particle level
the localized eigenstates of the electronic system and the spin
system
are identical.\cite{honecker_richter}
However,
for multi-particle states the different statistics and types of interaction
clearly become relevant.

\begin{figure}
\begin{center}
\includegraphics[width=\columnwidth]{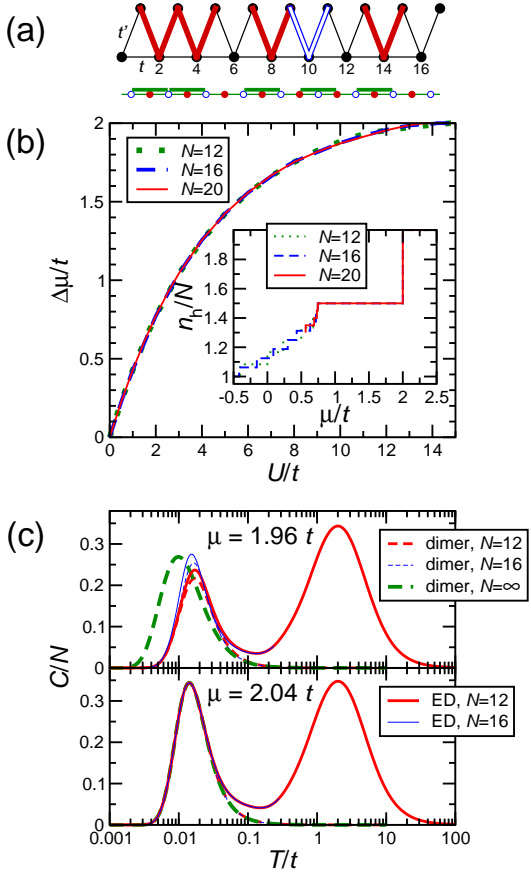}
\end{center}
\caption
{(Color online)
(a) Sawtooth lattice with auxiliary lattice for dimer mapping indicated below.
Each electron state localized in a valley of the sawtooth lattice corresponds
to a dimer on a site of the auxiliary lattice. Filled (open) symbols denote
spin-up (spin-down) electrons.
The presence of a dimer on a site of the auxiliary lattice
excludes the presence of dimers on the two adjacent sites.
(b) Main panel: Charge gap $\Delta \mu = E(N/2+1)-2\,E(N/2)+E(N/2-1)$
at quarter filling versus $U$.
Inset: Hole concentration $n_{\rm h}/N=2-n/N$ versus chemical potential $\mu$ for
$U=4\,t$.
(c) Specific heat $C$ per site $N$.
ED results are for $U=4\,t$.
Note that for $\mu=2.04\,t$ different system sizes are indistinguishable.}
\label{fig1}
\end{figure}

In the present Rapid Communication
we apply the ideas elaborated for frustrated AFM's having localized-magnon
GS's
to the Hubbard model with a flat band. We focus
on the sawtooth chain shown in Fig.~\ref{fig1}a
and characterize the complete manifold of highly degenerate GS's
for electron concentrations up to quarter filling.
Using a grand-canonical description and a mapping
to a classical hard-dimer problem, we derive
explicit analytical expressions
for the GS jump of the electron concentration, the residual
entropy at a particular value of the chemical potential $\mu_0$, and
the contribution of the GS manifold
to thermodynamic quantities such as the specific heat.
We believe
that the Hubbard model on other highly frustrated lattices shows
universal behavior as it has been demonstrated for spin systems.\cite{lm6,lm7}
Therefore,
we expect that similar considerations can be performed for
other lattices with flat bands, too.

The sawtooth-chain Hubbard model attracts much attention since the 90s
and was discussed earlier within different approaches.\cite{penc}
Moreover, a number of
compounds are known
to be described by the sawtooth-chain Heisenberg\cite{sawtooth} or periodic Anderson\cite{pam} models.

We consider the $N$-site Hubbard Hamiltonian
\begin{eqnarray}
H&=&\sum_{\sigma=\uparrow,\downarrow}H_{0\sigma}
 +H_U \, , \quad
 H_U = U\sum_{i}n_{i,\uparrow}n_{i,\downarrow} \, ,
\nonumber \\
H_{0\sigma}
&=& \sum_{\langle i,j \rangle}t_{i,j}
\left(c_{i,\sigma}^{\dagger}c_{j,\sigma}
+c_{j,\sigma}^{\dagger}c_{i,\sigma}\right)
  +\mu \sum_{i} n_{i,\sigma} \, ,
\label{01}
\end{eqnarray}
where
$i$ denotes the lattice sites (see Fig.~\ref{fig1}a),
$\langle i,j\rangle$ denote pairs of nearest neighbors,
the $c_{i,\sigma}^{\dagger}$ ($c_{i,\sigma}$)
are the usual fermion operators,
$n_{i,\sigma} = c_{i,\sigma}^{\dagger} c_{i,\sigma}$,
and periodic boundary conditions are implied.
We choose the relation $t^{\prime}=\sqrt{2}\,t > 0$ between
the hopping along the zig-zag path $t^{\prime}$
and the hopping along the base line $t$
in order to render the lowest single-electron band completely flat
(see e.g.\ Refs.~\onlinecite{tasaki,tasaki_ptp,honecker_richter}).
$U\ge 0$ is the on-site Coulomb repulsion
and $\mu$ is the chemical potential.
The sawtooth Hubbard model (\ref{01}) is a particular case of Tasaki's
model for which the GS exhibits saturated FM
for a half-filled flat band,
{\it i.e.}\ when the number of electrons is $n=N/2$.\cite{tasaki,tasaki_ptp}

{\it Localized-electron states. Eigenstates which do not feel $U>0$.}
Localized-electron states are a standard tool in the field of
flat-band FM (see, e.g.,
Refs.~\onlinecite{mielke,tasaki,tasaki_ptp,ichimura,nishino}).
As a basis for further discussion, we review the construction for
the sawtooth chain in a formulation inspired by
localized magnons in highly frustrated AFM's
\cite{lm1} (see also Ref.~\onlinecite{honecker_richter}).
Since the lowest single-electron band is completely flat,
it is possible to localize the corresponding states in real space,
namely on the three consecutive sites forming a `valley'
on the sawtooth chain. Let us introduce the operators
$l_{2j,\sigma}^{\dagger}
=c_{2j-1,\sigma}^{\dagger}-\sqrt{2}\,c_{2j,\sigma}^{\dagger}+c_{2j+1,\sigma}^{\dagger}$
which satisfy
$[H_{0\sigma},l^{\dagger}_{2j,\sigma}]_{-}=\varepsilon_{-}\, l^{\dagger}_{2j,\sigma}$
where $\varepsilon_{-}=-2\,t+\mu$ is the energy of the flat band.
Then the complete set of $N$
single-electron states belonging to the flat band
can be written as
 $l_{2j,\sigma}^{\dagger}\vert 0\rangle$
with $\vert 0\rangle$ denoting the vacuum state.
Application of $n$ distinct operators $l_{2j,\sigma}^{\dagger}$
to $\vert 0\rangle$ yields $n$-electron states. If the valleys belonging
to the $l_{2j,\sigma}^{\dagger}$  {\em with different spin} are disconnected
these are
exact eigenstates of the full Hamiltonian with energy $E_n = n\,\varepsilon_{-}$.
However, these simple product states do not exhaust the exact $n$-electron
eigenstates with energy $E_n$.

To construct {\em {all}}  $n$-electron states
with $U$-independent energy  $E_n = n\,\varepsilon_{-}$ systematically
we proceed as follows. We start with a fully spin-up-polarized state
$l_{2j,\uparrow}^{\dagger}l_{2(j+1),\uparrow}^{\dagger}\ldots l_{2(j+L-1),\uparrow}^{\dagger}\vert 0\rangle$,
{\it i.e.}\ localized electrons occupying a cluster of $L$ consecutive valleys.
Now we exploit
the $SU(2)$-invariance of the model (\ref{01}) to construct
new eigenstates by acting with
$S^-=\sum_i c_{i,\downarrow}^{\dagger}c_{i,\uparrow}$
on this state
and
using the relations
$[S^-,l_{2j,\uparrow}^{\dagger}]_{-}=l_{2j,\downarrow}^{\dagger}$
and
$[S^-,l_{2j,\downarrow}^{\dagger}]_{-}=0$.
$L$-fold application of $S^-$ yields the fully spin-down-polarized state
$l_{2j,\downarrow}^{\dagger}l_{2(j+1),\downarrow}^{\dagger}\ldots l_{2(j+L-1),\downarrow}^{\dagger}\vert0\rangle$.
However, the action of $(S^-)^m,\; m=1,\ldots,L-1,$
yields $L-1$ further eigenstates
with the same energy $L\,\varepsilon_{-}$ which cannot be expressed
by applying only one
product of $l_{2j,\sigma}^{\dagger}$ operators on $\vert 0\rangle$
but rather are given by linear combinations. Evidently, the same states are obtained
if we use $S_{{\rm cluster},L}^-
 =\sum_{i=2j-1}^{2(j+L)-1} c_{i,\downarrow}^{\dagger}c_{i,\uparrow}$
instead of $S^-$.
Finally, we can use these cluster
states as building blocks for product states containing
more than one cluster of occupied valleys.
As long as individual
clusters are separated by empty valleys, such product states
remain exact eigenstates.
All these states are eigenstates of
$H$ and have a definite value of total $S^z$, but they do not necessarily
carry a definite total spin $S$.

Let us discuss some further important properties of the above
constructed exact $n$-electron eigenstates with $n \le N/2$. 
Firstly, they are GS's for $U=0$. Note that their energy
$E_n=n\,\varepsilon_{-}$ is
independent of $U$.
Since $H_U$ is a positive semidefinite operator
for $U>0$ the on-site  $H_U$ can only increase energies. Thus,
the localized-electron states remain {\em GS's} for $U>0$.
Secondly, the localized $n$-electron states are {\em linearly independent},
which is connected with the fact
(as in the case of spin systems, see Ref.~\onlinecite{schmidt}) that
the middle site is unique to each valley.
Finally, we have to discuss
whether these states are the only GS's.
It is known from spin systems \cite{lm6,lm7} that a finite separation
of the flat one-particle band from the next dispersive band ensures
{\em completeness} of the localized states. In accordance with this,
the number of GS's obtained
by exact diagonalization (ED) of the Hubbard model for $n \le N/2$
agrees precisely with the number of localized $n$-electron states
$g_N(n)$ which we will compute next.

{\it Mapping to hard dimers.}
We show now that methods for
counting localized-magnon states in spin systems \cite{lm6a,lm6,lm7}
can be carried over to the Hubbard model. However, there is
one major difference between spin systems and electrons, namely
localized-magnon states of a spin system can be viewed
as hard-core bosons with nearest-neighbor intersite repulsion
whereas localized-electron states have to be considered
as a two-component (spin up, spin down)
fermionic system with on-site repulsion between different species.
The localized-magnon states in the $XXZ$ Heisenberg AFM
on the sawtooth chain
can be mapped onto hard dimers
on a simple chain with ${\cal N}=N/2$ sites.
\cite{lm6a,lm6}
A central result of this Rapid Communication is that a similar mapping to
hard dimers
exists also for the Hubbard model,
however, because of the presence of two components with twice as many sites
in the effective model, {\it i.e.}\ ${\cal N}=N$.

Above, we have constructed exact GS's as product states of
localized-electron states living in clusters. We now construct a one-to-one
correspondence of these localized-electron states and hard dimers.
First, we introduce an auxiliary 1D
lattice of $N/2$ cells
(see Fig.~\ref{fig1}a).
Each cell of the auxiliary lattice corresponds to one valley
of the sawtooth chain and contains two sites.
Each site of the auxiliary lattice can be occupied by a dimer
with a length which forbids simultaneous occupation of
neighboring sites by dimers (see Fig.~\ref{fig1}a).
A dimer on the left (right) site in a cell of the auxiliary
lattice corresponds to
a spin-up (spin-down) electron state trapped in a valley of the sawtooth
chain and is called a spin-up (spin-down) dimer.
Next we assign dimer configurations to the product states, considering
each
cluster of $L$ consecutive valleys which are occupied
by localized-electron states separately. We start from a cluster
with only spin-up electrons where the correspondence to dimers is
evident.
Then we consider the states obtained by repeated operation
of the cluster-operator $S_{{\rm cluster},L}^-$ on the spin-up-polarized state.
When one arrives at the state with only spin-down electrons, the
dimer description is again evident. At intermediate steps, application
of $(S_{{\rm cluster},L}^-)^m, \; m=1,\ldots,L-1,$
yields linear combinations of all possible
distributions (states) of $L-m$ spin-up and $m$ spin-down electrons on the cluster.
Since the coefficients of all these states  are
non-zero, we can choose the state where all spin-down electrons are located
at the right and all spin-up electrons at the left side of the cluster as
a representative for the whole
linear combination. For this choice of the
representative
the occupation of a right (spin-down) site
of the auxiliary lattice excludes occupation of its right neighbor
by a spin-up dimer.
On the other hand,
having a spin-down dimer as the right neighbor of a spin-up dimer
would correspond to two electrons being localized in the same valley
which is also excluded. We therefore arrive at the hard-dimer exclusion
rule that no two adjacent sites of the auxiliary lattice may be occupied
simultaneously.

Thus, we have established a one-to-one correspondence between product
localized $n$-electron states and  configurations of $n$ hard dimers.
Fig.~\ref{fig1}a
illustrates the mapping for a localized-electron state with $n=5$
electrons. It is important to note that there
is just one exception for periodic boundary conditions,
namely the cluster occupying the entire system, {\it i.e.}\
$n=N/2$ electrons (dimers).
In this case there are precisely two hard-dimer configurations
corresponding to the two spin-polarized (up and down) GS's.
However, these two states belong to a spin-$N/4$ $SU(2)$-multiplet,
{\it i.e.}\ the GS degeneracy is $g_N(N/2)=N/2+1$. The
fact that $N/2 - 1$ states are missing in the hard-dimer description can
be traced to a periodic cluster having no right boundary.
Hence, we conclude that the degeneracy of the GS $g_N(n)$
in the subspaces $n=0,1,\ldots,N/2-1$
equals the canonical partition function of hard dimers ${\cal{Z}}(n,N)$,
$g_N(n)={\cal{Z}}(n,N)$, and $g_N(N/2)={\cal{Z}}(N/2,N)+N/2-1$.

{\it Thermodynamics.}
Due to their huge degeneracy, the GS's in the sectors with $n \le N/2$
will dominate the grand-canonical
partition function of the Hubbard model (\ref{01}) at
low temperatures and for a chemical potential close to $\mu_0=2\,t$.
We can use the mapping to hard dimers to calculate this contribution
exactly.
Noting that the energy of a localized $n$-electron state is
$E_n=n\,\varepsilon_{-}$ we can write the grand-canonical partition function
for localized-electron states as
$\Xi(\beta,\mu,N)
=\sum_{n=0}^{N/2}g_N(n)\exp(-\beta n\varepsilon_{-})$.
The r.h.s.\ of this equation contains the grand-canonical partition
function of hard dimers
on a chain of $N$ sites,
which can be computed using the transfer-matrix approach.
This leads to
$\Xi(\beta,\mu,N)=\lambda_+^N+\lambda_-^N
+ (N/2-1) \exp(N\,x/2)$,
$\lambda_{\pm}=1/2\pm\sqrt{1/4+\exp x}$.
Note that only the combination $x=\beta(2\,t-\mu)$
enters all thermodynamic quantities in the hard-dimer description.
In the thermodynamic limit $\Xi(\beta,\mu,N)$
is identical to the behavior of 1D
hard dimers,
since the sector with $n=N/2$ (not described by hard dimers) becomes irrelevant.
For $N \to \infty$, the
thermodynamic potential becomes
$-\beta\Omega(\beta,\mu,N)/N=\ln\lambda_{+}$
leading to simple expressions for thermodynamic
quantities \cite{lm6a,lm6,lm7} such as
\begin{eqnarray}
\label{02}
\frac{C(\beta,\mu,N)}{N}
=
\frac{\left(\beta(2\,t-\mu)\right)^2\exp(\beta(2\,t-\mu))}{8\left(\frac{1}{4}+\exp(\beta(2\,t-\mu))\right)^{\frac{3}{2}}}
\end{eqnarray}
for the specific heat. In particular, at $\mu=\mu_0=2\,t$ we have
$\varepsilon_{-} = 0$ resulting in
a finite residual entropy
$S/N=\ln((1+\sqrt{5})/2) = 0.48121\ldots$.

{\it Numerical results.}
In order to estimate the range of validity of the hard-dimer description
for the
Hubbard model (\ref{01}), we perform complementary
numerical computations for finite systems:
(i) GS properties are computed for the Hubbard model
 with $N\le 20$ using the Lanczos method.
(ii)
Thermodynamic quantities are derived from full diagonalization
of the Hubbard model with $N \le 12$ using
 a custom shared memory parallelized Householder algorithm
to diagonalize complex matrices of
dimension up to 121~968.

The inset of Fig.~\ref{fig1}b shows
the hole concentration $n_{\rm h}/N=2-n/N$ versus $\mu$.
Like for spin systems,\cite{lm1}
the main characteristics are a size-independent jump of $n/N$
from $0$ to $1/2$ and a plateau at $n/N=1/2$.
This plateau determines the range of validity
of the hard-dimer picture at $T=0$.
The main panel of Fig.~\ref{fig1}b presents the
plateau width, {\it i.e.}\ the charge gap,
versus $U$ for $N=12$, $16$, and $20$ and one
observes almost no finite-size dependence. Since the charge gap
vanishes at $U=0$ we infer that its appearance is due to on-site repulsion.

As an example for thermodynamic properties we consider the specific
heat $C$ shown in Fig.~\ref{fig1}c for $\mu = 2.04\,t$
and $1.96\,t$. For the Hubbard model (shown here
by solid lines for $U=4\,t$) $C$ exhibits two maxima. First, there is
a high-$T$
maximum around $T \approx 2\,t$, like for any
system with a finite bandwidth. Second, there is a low-$T$
maximum which is located at $T$ of the order $t/100$ for
$|\mu-\mu_0|=0.04\,t$. This low-$T$
maximum is due to
the manifold of localized-electron GS's. Indeed,
the hard-dimer results (shown by dashed lines in Fig.~\ref{fig1}c)
are indistinguishable from the full Hubbard model in the region
of the low-$T$
maximum at $\mu = 2.04\,t$. For
$\mu=1.96\,t$ we observe
deviations for a fixed $N$ even at
low $T$ which can be attributed to
excited states
in the Hubbard model. Nevertheless, also this low-$T$ maximum
is qualitatively described by hard dimers and better agreement can
be obtained by considering larger values of $U$.  The low-$T$ maximum
shifts to lower temperatures for $\mu \to \mu_0 = 2\,t$ and disappears
at $\mu=\mu_0$ in favor of a macroscopic GS degeneracy.
Note that in the
hard-dimer picture the thermodynamic limit can be carried out explicitly,
see (\ref{02}) for the specific heat.

{\it Conclusions.}
In summary,
we have given an exact solution for the GS properties of a
correlated many-electron system
in a certain range of the chemical potential
and studied the low-$T$ thermodynamics.
Although we focus here on a specific lattice (the sawtooth chain),
the Hubbard model on other highly frustrated lattices
should exhibit qualitatively similar behavior:
firstly, this has been demonstrated for spin systems;\cite{lm6,lm7}
secondly, preliminary calculations (not shown here)
for other one-dimensional lattices
yield similar results;
thirdly, a macroscopic GS degeneracy
in a general flat-band Hubbard model
can be derived from a trivial lower bound
\cite{honecker_richter}
(for the kagome lattice, this
also follows from early work \cite{mielke}).
For the sawtooth chain, a mapping to hard dimers yields the degeneracy
of the exact many-electron GS's
and their contribution to the thermodynamics.
We have discussed the specific heat $C$
and observed the emergence of a low-$T$ maximum
for $\mu \ne \mu_0$ (Fig.~\ref{fig1}c)
which is at least qualitatively described by hard dimers.
Since this low-$T$ maximum in $C$ is related to the
macroscopic GS degeneracy at $\mu = \mu_0$,
such a low-$T$ maximum in $C$ should be a characteristic fingerprint
of a flat-band FM which is also accessible experimentally.
Note that the derivation of exact eigenstates is valid only for $t^{\prime}=\sqrt{2}\,t$.
Deviations from this relation 
will lift the macroscopic GS degeneracy at $\mu = \mu_0$,
but the ferromagnetic GS for $\mu < \mu_0$
is protected \cite{penc,24}
by the presence of a charge gap (Fig.~\ref{fig1}b).
Preliminary calculations (not shown here) demonstrate
that the low-$T$ thermodynamic behavior remains qualitatively unchanged
for small deviations from the ideal geometry,
like in previous studies of localized magnons in spin systems.\cite{lm6}

The localized-electron picture can also be used to study magnetic
properties of the corresponding Hubbard models.
Indeed, the sawtooth model exhibits fully polarized GS's for the sectors
$n=N/2$ \cite{tasaki,tasaki_ptp} and $n=N/2 - 1$, whereas
the average over all degenerate GS's yields
exactly $3/5$ of the maximal polarization
in the sector with $n=N/2-2$ and for $N \to \infty$.
However, we can show (details will be presented elsewhere)
that there is no finite range of FM in
the sawtooth Hubbard model
for electron concentrations $n/N < 1/2$ as $N \to \infty$.

{\it Acknowledgments.}
The present study was supported by the DFG (Project HO~2325/4-1),
the Rechenzentrum of the TU Braunschweig and the HLRN Hannover,
as well as the MPIPKS-Dresden.
The authors are indebted to H.~Frahm and W.~Brenig
who brought flat-band FM to their attention.


\begin{thebibliography}{999}

\bibitem{thehubbard}
{\it{The Hubbard Model --- A Reprint Volume}},
edited by
A.~Montorsi
(World Scientific, Singapore, 1992).

\bibitem{lieb}
E.~H.~Lieb,
arXiv:cond-mat/9311033.

\bibitem{tasaki_jpcm}
H.~Tasaki,
J. Phys.: Condens. Matter {\bf 10}, 4353 (1998).

\bibitem{vollhardt}
D.~Vollhardt, N.~Bl\"{u}mer, K.~Held, M.~Kollar, J.~Schlipf, and M.~Ulmke,
Z. Phys. B {\bf 103}, 283 (1997).

\bibitem{mielke}
A.~Mielke,
J. Phys. A {\bf 24}, L73 (1991);
%A.~Mielke,
ibid. %J. Phys. A
{\bf 24}, 3311 (1991);
%A.~Mielke,
ibid. %J. Phys. A
{\bf 25}, 4335 (1992).

\bibitem{tasaki}
H.~Tasaki,
Phys. Rev. Lett. {\bf 69}, 1608 (1992);
A.~Mielke and H.~Tasaki,
Commun. Math. Phys. {\bf 158}, 341 (1993).

\bibitem{tasaki_ptp}
H.~Tasaki,
Prog. Theor. Phys. {\bf 99}, 489 (1998).

\bibitem{ichimura}
M.~Ichimura, K.~Kusakabe, S.~Watanabe, and T.~Onogi,
Phys. Rev. B {\bf 58}, 9595 (1998).

\bibitem{tanaka}
A.~Tanaka and H.~Ueda,
Phys. Rev. Lett. {\bf 90}, 067204 (2003).

\bibitem{sekizawa}
T.~Sekizawa,
J. Phys. A: Math. Gen. {\bf 36}, 10451 (2003).

\bibitem{nishino}
S.~Nishino, M.~Goda, and K.~Kusakabe,
J. Phys. Soc. Jpn. {\bf 72}, 2015 (2003);
S.~Nishino and M.~Goda,
ibid. {\bf 74}, 393 (2005).

\bibitem{tanaka_tasaki}
A.~Tanaka and H.~Tasaki,
Phys. Rev. Lett. {\bf 98}, 116402 (2007).

\bibitem{lm1}
J.~Schnack, H.-J.~Schmidt, J.~Richter, and  J.~Schulenburg,
Eur. Phys. J. B {\bf 24}, 475 (2001);
J.~Schulenburg, A.~Honecker, J.~Schnack, J.~Richter, and H.-J.~Schmidt,
Phys. Rev. Lett. {\bf 88}, 167207 (2002);
J.~Richter, J.~Schulenburg, A.~Honecker, J.~Schnack, and H.-J.~Schmidt,
J. Phys.: Condens. Matter {\bf 16}, S779 (2004).

\bibitem{lm5}
J.~Richter, O.~Derzhko, and J.~Schulenburg,
Phys. Rev. Lett. {\bf 93}, 107206 (2004).

\bibitem{lm6a}
M.~E.~Zhitomirsky and A.~Honecker,
J. Stat. Mech.: Theor. Exp. P07012 (2004).

\bibitem{lm6}
O.~Derzhko and J.~Richter,
Phys. Rev. B {\bf 70}, 104415 (2004);
Eur. Phys. J. B {\bf 52}, 23 (2006).

\bibitem{lm7}
M.~E.~Zhitomirsky and H.~Tsunetsugu,
Phys. Rev. B {\bf 70}, 100403(R) (2004);
Prog. Theor. Phys. Suppl. {\bf 160}, 361 (2005);
J.~Richter, O.~Derzhko, and T.~Krokhmalskii,
Phys. Rev. B {\bf 74}, 144430 (2006).

\bibitem{honecker_richter}
A.~Honecker and J.~Richter,
Condensed Matter Physics (L'viv) {\bf 8}, 813 (2005).

\bibitem{penc}
K.~Penc, H.~Shiba, F.~Mila, and T.~Tsukagoshi,
Phys. Rev. B {\bf 54}, 4056 (1996);
H.~Sakamoto and K.~Kubo, J. Phys. Soc. Jpn. {\bf 65}, 3732 (1996);
Y.~Watanabe and S.~Miyashita, J. Phys. Soc. Jpn. {\bf 66}, 2123 and 3981
(1997).

\bibitem{sawtooth}
G.~C.~Lau, B.~G.~Ueland, R.~S.~Freitas, M.~L.~Dahlberg, P.~Schiffer, and R.~J.~Cava,
Phys. Rev. B {\bf 73}, 012413 (2006).

\bibitem{pam}
H.~N.~Kono and Y.~Kuramoto,
J. Phys. Soc. Jpn. {\bf 75}, 084706 (2006).

\bibitem{schmidt}
H.-J.~Schmidt, J.~Richter, and R.~Moessner,
J. Phys. A {\bf 39}, 10673 (2006).

\bibitem{24}
H.~Tasaki, 
Phys. Rev. Lett. {\bf 75}, 4678 (1995);
A.~Mielke, 
Phys. Rev. Lett. {\bf 82}, 4312 (1999);
J. Phys. A {\bf 32}, 8411 (1999).

\end{thebibliography}
\end{document}